\documentstyle[preprint,prd,aps,epsfig,floats]{revtex}
\newcommand{\be}{\begin{equation}}
\newcommand{\ee}{\end{equation}}
\newcommand{\bea}{\begin{eqnarray}}
\newcommand{\eea}{\end{eqnarray}}

\tightenlines
\begin{document}
\firstfigfalse
\title{Closed-Time Path Integral Formalism and
Medium Effects of Non-Equilibrium QCD Matter}

\author{Chung-Wen Kao\thanks{Present Address: Theoretical Physics Group,
Department of Physics and Astronomy, University of Manchester,Manchester, M139PL, UK}, 
Gouranga C. Nayak\thanks{Present Address: T-8, Theoretical Division,
Los Alamos National Laboratory, Los Alamos, NM 87545, USA}, 
and Walter Greiner}

\address
{{\small\it{Institut f\"ur Theoretische Physik,
J. W. Goethe-Universit\"at,
60054 Frankfurt am Main, Germany}}}
\maketitle

\begin{abstract} 
We apply the closed-time path integral formalism to
study the  medium effects of non-equilibrium gluon matter. We derive the
medium modified resummed gluon propagator to the one loop level
in non-equilibrium in the covariant gauge. The gluon propagator we derive 
can be used to remove the infrared divergences in the secondary parton
collisions to study thermalization of minijet parton plasma at RHIC and
LHC.

\end{abstract}  
\bigskip

   
\section{introduction}
Much effort has been given to detect a new form of
matter known as quark-gluon plasma (QGP) for decades. 
The relativistic heavy-ion collider
experiments at RHIC (Au-Au collisions at $\sqrt{s}$ = 200 GeV) 
and LHC (Pb-Pb collisions at $\sqrt{s}$ = 
5.5 TeV) will provide the best opportunity to
study such a state of matter in the 
laboratory. Perturbative QCD estimates
that the energy density of produced jets and minijets might be larger 
than 50 and 1000 GeV/$fm^3$ at RHIC and LHC 
\cite{nayak,keijo1} which is much larger than the 
required energy density
to produce quark-gluon plasma.
As QGP lives for a very short time (several fermi)
in a small volume ($\sim$ 100 fm$^3$) a direct detection of this phase
is not possible. Hence various indirect signatures are proposed for its
detection. The prominent among them are: 1) $J/{\Psi}$ suppression \cite{satz},
2) strangeness enhancement \cite{str}, dilepton and direct photon
production \cite{dil,ph}. However, many uncertainties
exist which make it difficult to claim the existence of the
quark-gluon plasma. The main uncertainty lies in the
lack of an accurate determination
of the space-time evolution of the quarks and gluons produced just after
the collision of two nuclei at RHIC and LHC. 
While the quark-gluon plasma
during the equilibrium stage is described by Bjorken's hydrodynamic 
evolution equations, it is much more difficult
to determine the space-time evolution of partons in the 
pre-equilibrium stage. An accurate study of the pre-equilibrium stage
will determine the equilibration time and initial conditions
for hydrodynamic evolutions in the equilibrium stage. This study is also very
important for the accurate determination of various signatures of the quark-gluon
plasma. 

To describe the space-time evolution of quark-gluon plasma at RHIC and LHC
one needs to know how the partons are formed in these high energy
nuclear collisions. The hard parton (jets and minijets) 
production can be calculated by using pQCD. 
The pre-equilibrium evolution of these hard partons can be studied
by solving relativistic transport equations with secondary collisions
among these partons taken into account 
\cite{nayak,geiger,wang12,wong,gyulassy}.
However, soft parton production can not be computed within pQCD formalism.
There are coherent effects for the soft partons and they may be
described by formation of a classical chromofield 
\cite{bhal,naya,others,denis,larry1,roberts,bla}.
For simplicity, we will only consider the evolution of minijet plasma 
which can be studied by solving relativistic transport equation:
\be
p^{\mu}\partial_{\mu}f(x,p)~=~C(x,p),
\label{trs1}
\ee
with secondary collision among the partons taken into account. In the 
above equation
\bea
&&C(x,p)~=~ \int \frac{d^3p_2}{(2\pi)^3p_2^0}~ \int \frac{d^3p_3}{(2\pi)^3p_3^0}
~\int \frac{d^3p_4}{(2\pi)^3p_4^0}~ |M(pp_2\rightarrow p_3p_4)|^2 
\delta^4(p+p_2-p_3-p_4) \nonumber \\
&&~[f(x,p_3)f(x,p_4)(1+f(x,p))(1+f(x,p_2))-f(x,p)f(x,p_2)(1+f(x,p_{3}))(1+f(x,p_{4}))]
\label{col1}
\eea
is the collision term for a partonic 
scattering process $pp_2\rightarrow p_3p_4$. In the above
expression $p$, $p_2$, and $p_3$, $p_4$ are the four momentum of the partons
before and after the collision. Different partonic 
scattering processes which can be considered are like 
$gg \rightarrow gg$, $qq\rightarrow qq$, $q\bar q \rightarrow q 
\bar q$ and $gq \rightarrow gq$ etc.. 
As the gluons are the
dominant part of the minijet production we only consider 
$gg \rightarrow gg$ process in the following.
The squared matrix element for this process is given by:
\be
|M(\hat{s},\hat{u},\hat{t})|^2 ~=~\frac{9\pi^2\alpha_s^2}{8}~[3
-\frac{\hat{u}\hat{t}}{\hat{s}^2}
-\frac{\hat{u}\hat{s}}{\hat{t}^2}
-\frac{\hat{s}\hat{t}}{\hat{u}^2}],
\label{mat}
\ee
where
$\hat{s}~=~(p+p_2)^2~=~(p_3+p_4)^2$, $\hat{t}~=~(p-p_3)^2~=~(p_2-p_4)^2$,
$\hat{u}~=~(p-p_4)^2~=~(p_2-p_3)^2$ are the Mandelstam variables. For massless gluon they are related by
\be
\hat{t}=-\frac{\hat{s}}{2}~[1-cos\theta], \,\,\,
\hat{u}=-\frac{\hat{s}}{2}~[1+cos\theta],  
\label{tu}
\ee
where $\theta$ is the center of mass scattering angle which goes from
$0 \rightarrow \frac{\pi}{2}$ for identical partons in the final state. 
When one puts
$|M(\hat{s},\hat{u},\hat{t})|^2$ from Eq. (\ref{mat}) in Eq. (\ref{col1})
one encounters divergence in the collision term at small angle
($\theta \rightarrow$ 0) or small momentum transfer $\hat{t} \rightarrow$ 0.
This divergence is inevitable as long as we use the free propagator in vaccuum to
evaulate the Feynman diagrams.
However, this infrared divergence can be
removed when one uses medium modified propagators instead of the
vacuum propagator to evaluate the collision term in medium 
\cite{art,naka,bll2,wang12,ruppert}.
The medium modified propagtaors
have been obtained in the thermal field theory for the case of an 
equilibrium plasma.
However, these finite temperature calculations 
are valid only in equilibrium where there is a static temperature such
as in the case of a heat bath and the system is isotropic. 
For realistic situations in the 
high energy heavy-ion collisions at RHIC and LHC
the partons formed at the initial time are 
in non-equilibrium and finite temperature QCD
is not applicable at this stage. One has to use
closed-time formalism to compute various quantities in 
non-equilibrium \cite{sch,mah,otrs,hu}. 
In this paper, using closed-time path integral formalism,
we derive the medium modified resummed gluon propagator in non-equilibrium
to the one loop order in the covariant
gauge which is necessary to obtain finite collision term
to study equilibration of expanding
minijet plasma at RHIC and LHC.

The paper is organized as follows. In section II we describe the formulation
of closed-time path integral of SU(3) guage theory. In section III we
derive the resummed gluon propagator to one loop level.
We summarize and conclude our main results in section IV.

\section{Closed-Time Path Integral Formalism in Gauge Theory}

We consider
SU(3) pure gauge theory which is QCD without quarks.
In high energy heavy-ion collisions at RHIC and LHC most of the
parton formed are gluons. Hence we concentrate on gluons only. However,
extending close-time path integral formalism to quarks is straightforward.
As the two nuclei travel almost at a speed of light at RHIC and LHC
the system is dynamically evolving and many quantities
have to be formulated in Boost invariant way \cite{bjorken}. For this expanding
system of partons we work in the covariant gauge in this paper. 
The QCD action without quark is given by:
\be
S=\int d^4x [-\frac{1}{4}F^{a\mu \nu}F^a_{\mu \nu}~-~\frac{1}{2\xi}(G^a)^2
-{\cal L}_{FP} ],
\label{gaugac}
\ee
where the gluon field tensor:
\be
F^{a\mu \nu}~=~\partial^{\mu}A^{a\nu}~-~\partial^{\nu}A^{a\mu}~+~gf^{abc}
A^{b\mu}A^{c\nu},
\ee
the ghost Lagrangian density: 
\be
{\cal L}_{FP}~=~[\partial_{\mu}\bar C] D^{\mu}[A] C
\ee
and the gauge fixing term given by
\be
G^a~=~\partial_{\mu}A^{a\mu}.
\ee
In Eq. (\ref{gaugac}) $\xi$ is the gauge fixing parameter in covariant gauge.
In the closed-time path integral formalism the gauge field $A$, 
the ghost field $C$ and the corresponding
sources $j, \chi$ are defined in both the time branches. 
To make the formulas simpler
we denote the fields and the corresponding sources by following
common notations:
\be
Q=~(A,C, \bar C), ~~~~~ ~~~~~~J~=(j,\bar{\chi}, \chi).
\ee
Denoting $Q^+$, $Q^-$ and $J^+$,$J^-$ the fields and the sources
on the upper and lower branch of the time path, 
the in-in generating functional becomes:
\be
Z[J_+,J_-,\rho]~= ~\int DQ^r~<Q^+,t_0|~\rho~|Q^-,t_0>~e^{i(S[Q^r]~+~J_r \cdot Q^r)},
\ee
where $S[Q^r]~=~S[Q^+]-S^{*}[Q^-]$ with $r=+,-$. More explicitly,
the above equation can be written as:
\bea
Z[J_+,J_-,\rho]~= ~\int [dQ^+]d[Q^-]~<Q^+,t_0|~\rho~|Q^-,t_0>~e^{i[
S_0[Q] ~+~S_{int}[{\bf Q}]~+~
Tr{\bf J \cdot Q}]},
\label{z0p}
\eea
where 
\bea
S_0[Q]
~=~\sum_{r,s=+,-}~\int \frac{d^4p}{(2\pi)^4}~[\frac{1}{2}A^r_{\mu}(p)[
G_{\mu \nu}^{-1}(p)]^{rs}A^s_{\nu}(-p)~+~{\bar C}^r(p)[S^{-1}(p)]^{rs}C^s(-p)],
\label{s0p}
\eea
and
\be
Tr{\bf J \cdot Q}~=~\int \frac{d^4p}{(2\pi)^4}[j^+_{\mu}(-p)A^+_{\mu}(p)
~+~{\bar C}^+(-p) \chi^+(p)~+~{\bar \chi}^+(-p) C^+(p)~+~(+ \rightarrow -)~].
\ee
In the Eq. (\ref{s0p})
$G_{\mu \nu}(p)$ and $S(p)$ are gluon and ghost free propagators 
and in Eq. (\ref{z0p}) $S_{int}[{\bf Q}]~=~S_{int}[Q^+]-S^{*}_{int}[Q^-]$.

In covariant gauge the further complications arise 
because of the presence of the unphysical ghost fields. 
In equilibrium one can define the ghost distribution function (BE)
and hence can deal with ghost fields in the medium
even in covariant gauge. However, in non-equilibrium
there is no easy procedure to obtain a ghost distribution
function in the QCD medium at RHIC and LHC. Quark and gluon 
distribution functions
in non-equilibrium situations at RHIC and LHC can be obtained
from minijets by using pQCD or via other methods
\cite{geiger,wang12,wong,gyulassy,bhal,nayak,fred,muller}.
For this reason we work
in the Landshoff and Rebhan scheme of frozen ghost formalism
\cite{land,petro}, where
the gauge theory in the
covariant gauge is obtained by restricting the space of initial
state $|Q^r,t_0>$ to the physical one. This means gluons with
spatially transverse polarization will contribute to the trace. Hence
the non-local Kernel $K$ appearing in the path integral of the
generating function \cite{hu}
will couple only to the transverse component 
of the gauge field $A$. 

Now we consider a cylindrically symmetric 
expanding system of partons in 1$\oplus$1 dimension.  
For this purpose we introduce the flow velocity of the medium
\be
u^{\mu}~=~(\cosh\eta,0,0,\sinh\eta),
\label{flow}
\ee
where $\eta=\frac{1}{2}\ln{\frac{t+z}{t-z}}$
is the space-time rapidity and $u_{\mu}u^{\mu}~=~1$. We define
the four symmetric tensors \cite{pr,wel,land}:
\begin{eqnarray}
T_{\mu\nu}(p)&=&g_{\mu\nu}-\frac{(u\cdot p)(u_{\mu}p_{\nu}
+u_{\nu}p_{\mu})-p_{\mu}p_{\nu}-p^{2}u_{\mu}u_{\nu}}
{(u\cdot p)^{2}-p^{2}}, \nonumber \\
L_{\mu\nu}(p)&=&\frac{-p^{2}}{(u\cdot p)^{2}-p^{2}}
\left(u_{\mu}-\frac{(u\cdot p)p_{\mu}}{p^{2}}\right)
\left(u_{\nu}-\frac{(u\cdot p)p_{\nu}}{p^{2}}\right), \nonumber \\
C_{\mu\nu}(p)&=&\frac{1}{\sqrt{2[(u\cdot p)^{2}-p^{2}]}}\left[\left(u_{\mu}
-\frac{(u\cdot p)p_{\mu}}{p^{2}}\right)p_{\nu}+\left(u_{\nu}
-\frac{(u\cdot p)p_{\nu}}{p^{2}}\right)p_{\mu}\right] {\rm and} \nonumber \\
D_{\mu\nu}(p)&=&\frac{p_{\mu}p_{\nu}}{p^{2}},
\label{tmunu}
\end{eqnarray}
which are required for a cylindrically symmetric system in 1$\oplus$1
expanding plasma at RHIC and LHC in the very early stage.

Here $T^{\mu\nu}$ is transverse with respect to the flow-velocity but
$L^{\mu\nu}$ and $D^{\mu\nu}$ are mixtures of space-like and time-like
components. These tensors satisfy the following transversality properties
with respect to $p^{\mu}$:
\be
p_{\mu}T^{\mu\nu}(p)~=~p_{\mu}L^{\mu\nu}(p)~=~0,~~~~~~~~~~~~
p_{\mu}p_{\nu}C^{\mu\nu}(p)~=~0.
\ee
In addition to this, the above tensors satisfy the following properties:
\begin{equation}
\begin{array}[c]{c}
T\cdot L=T\cdot C=T\cdot D=0,\,\,\,
T+L+D=1,\\
T\cdot T=T,\,\,L\cdot L=L,\,\,C\cdot C=\frac{1}{2}(L+D),\\
Tr C\cdot L=Tr C\cdot D =0.
\end{array}
\end{equation}
Any symmetric tensor $S^{\mu\nu}$ can be written in terms of the 
above four tensors:
\be
S^{\mu\nu}~=~ a ~T^{\mu\nu}+ b ~L^{\mu\nu}+ c ~C^{\mu\nu}+ d ~D^{\mu\nu}
\ee
with
\be
a=\frac{1}{2}Tr~T\cdot S, ~~b~=~ Tr~L\cdot S, ~~c~=~ -Tr~C\cdot S,
 ~~d~=~ Tr~D\cdot S.
\ee

In terms of this tensor basis the gluon propagator in the covariant
gauge is given by:
\begin{eqnarray}
G_{\mu\nu}(p)_{ij}&=&-iT_{\mu\nu}(p)\left(\left[G(p)\right]^{vac}_{ij}
~+~\left[G(p)\right]^{med}_{ij}\right)
-i(L_{\mu\nu}(p)
+\xi D_{\mu\nu}(p))\left[G(p)\right]^{vac}_{ij} \nonumber\\
&=&-i\left(g_{\mu\nu}+(\xi-1)D_{\mu\nu}(p)\right)\left[G(p)\right]^{vac}_{ij}
-iT_{\mu\nu}(p)\left[G(p)\right]^{med}_{ij},
\end{eqnarray}
where $i,j=+,-$. The forms of $G(p)_{ij}$ are
\begin{equation}
\left[G(p)\right]^{vacuum}_{ij}=\left(\begin{array}{cc}
\frac{1}{p^2+i\epsilon}& 0 \\
0 &\frac{-1}{p^2-i\epsilon}
\end{array}
\right)-2\pi i\delta(p^2)\left(\begin{array}{cc}
0&\theta(-p_{0})\\
\theta(p_{0})&0
\end{array}
\right),
\end{equation}
and
\begin{equation}
\left[G(p)\right]^{medium}_{ij}=-2\pi i\delta(p^2)\tilde{f}(\vec{p})\left(\begin{array}{cc}
1&1\\
1&1
\end{array}
\right),\,\,\,\tilde{f}(\vec{p})=f(\vec{p})\cdot\theta(p_{0})+f(-\vec{p})
\cdot\theta(-p_{0}).
\end{equation}
Here $f(\vec{p})$ is the distribution function.
Note that in this Landshoff-Rehban scheme
the transverse component of the gauge
propagator proportional to $T^{\mu\nu}$ contains the medium effect
which is gauge-parameter independent.
As the initial density matrix contains only sum over transverse
polarizations of gluons the ghost propagator is simply given by: 
\begin{equation}
S(p)_{ij}=-i\left[G(p)\right]^{vac}_{ij},
\end{equation}
because ghost fields do not couple to the kernel $K$ in the generating
function.

\section{Resummed Gluon Propagator at One Loop Level in Non-Equilibrium}

In the last section we derived the gauge field propagator in the
medium for the free part of the action $S_0$. In this section
we will consider the interaction term of the action $S_{int}$. 
For such a situation one will have to consider the full
propagator instead of the free propagator in the medium. This
implies one has to solve the full Schwinger-Dyson equation which
is practically impossible. Usually one has to truncate the series 
at one or two loop level of the self-energy. 
Such a truncation demands that the QCD coupling constant should not
be too large. We consider here the evolution of the 
gluon-minijet plasma at RHIC and LHC
where the average transverse momentum of the partons are
found to be large at the earlry stage \cite{nayak}. Therefore the coupling
constant corresponding to such average transverse momentum is 
found to be small.
In this situation we truncate the Schwinger-Dyson equation
at one loop level of the self energy
and consider the corresponding resummed two-point Green's function
$\tilde{G}$.

The resummed two-point Green's function $\tilde{G}$ of gluon
can be decomposed as:
\begin{equation}
\tilde{G}_{\mu\nu}(p)=-iT_{\mu\nu}(p)\tilde{G}^{T}(p)
-iL_{\mu\nu}(p)\tilde{G}^{L}(p)
-i\xi D_{\mu\nu}(p)\tilde{G}^{D}(p),
\end{equation}
where $\tilde{G}^T$, $\tilde{G}^L$, $\tilde{G}^D$ correspond to $T$, $L$ and
$D$ components respectively. The last part $\tilde{G}^D(p)$ is identical
to the vacuum part \cite{pr} and hence we do not consider it any more.
It can be mentioned that there are separate
Dyson-Schwinger equation for different components which does not couple with
each other. 
 
Dyson-Schwinger equations for different components can be written as 
\begin{equation}
\left[\tilde{G}^{T,L}(p)\right]_{ij}=\left[G^{T,L}(p)\right]_{ij}
+\sum_{l,k}\left[G^{T,L}(p)\right]_{il}
\cdot\left[\Pi^{T,L}(p)\right]_{lk}\cdot \left[\tilde{G}^{T,L}(p)\right]_{kj}.
\end{equation}
Please remember that $i,j,k,l~=~+,-$ and suppression of Lorentz and color
indices in the above equation is understood.

Instead of using the matrix form of this equation, 
we prefer to use the retarded, advanced and symmetric
Green's functions. The retarded and advanced resummed Green's function
are found to be:
\begin{equation}
\tilde{G}^{T,L}_{R,A}(p)=G^{T,L}_{R,A}(p)+G^{T,L}_{R,A}(p)
\cdot\Pi^{T,L}_{R,A}(p)\cdot \tilde{G}^{T,L}_{R,A}(p).
\end{equation}

The straightforward solution of the above equation gives:
\begin{equation}
\tilde{G}^{T,L}_{R,A}(p)=\frac{G^{T,L}_{R,A}(p)}{1-G^{T,L}_{R,A}(p)\cdot
\Pi^{T,L}_{R,A}(p)}=\frac{1}{p^2-\Pi^{T,L}_{R,A}(p) \pm i sgn(p_{0})\epsilon},
\end{equation}
where the self-energy contains the medium effects.
Similar but more complicated equation is obtained for the resummed
symmetric Green's function:
\begin{equation}
\begin{array}[c]{c}
\tilde{G}^{T,L}_{S}(p)=G^{T,L}_{S}(p)+G^{T,L}_{R}(p)\cdot\Pi^{T,L}_{R}(p)
\cdot \tilde{G}^{T,L}_{S}(p)\\
+G^{T,L}_{S}(p)\cdot\Pi^{T,L}_{A}(p)\cdot \tilde{G}^{T,L}_{A}(p)
+G^{T,L}_{R}(p)\cdot\Pi^{T,L}_{S}(p)\cdot \tilde{G}^{T,L}_{A}(p).
\end{array}
\end{equation}
After some algebra it can be shown that
\begin{equation}
\begin{array}[c]{c}
\tilde{G}^{T,L}_{S}(p)=\left[1+2f(\vec{p})\right]sgn(p_{0})[\tilde{G}^{T,L}_{R}(p)
-\tilde{G}^{T,L}_{A}(p)]\\
+\left(\Pi_{S}^{T,L}(p)-(1+2f(\vec{p}))sgn(p_{0})[\Pi^{T,L}_{R}(p)-\Pi^{T,L}_{A}(p)]\right)
\times \tilde{G}^{T,L}_{R}(p)\times \tilde{G}^{T,L}_{A}(p).
\end{array}
\end{equation}

Let us evaluate the various components of the self energy. For short
hand notation we denote {\bf Gl} for gluon loop, {\bf Ta} for tapole diagram
(see Fig. 1) and {\bf FP} represents ghost loop diagram. 
Explicitly the expressions for various self energies are given by:
\begin{eqnarray}
\left[\Pi^{ab,}_{Gl;\mu\nu}(p)\right]_{kl}&=&(k\times l) \frac{g^{2}}{2i}\delta_{ab}N_{c}
\cdot
\int\frac{d^{4}q}{(2\pi)^{4}}\left[G^{\alpha\gamma}(q)\right]_{kl}
\left[G^{\beta\delta}(p-q)\right]_{kl}  \nonumber\\
&\cdot &\left[g_{\mu\alpha}(-p-q)_{\beta}+
g_{\alpha\beta}(2q-p)_{\mu}+g_{\beta\mu}(2p-q)_{\alpha}\right]
 \nonumber\\
&\cdot&\left[g_{\nu\gamma}(p+q)_{\delta}+
g_{\gamma\delta}(p-2q)_{\nu}+g_{\nu\delta}(q-2p)_{\gamma}\right].
\nonumber \\
\left[\Pi^{ab}_{Ta;\mu\nu}(p)\right]_{kl}&=&-g^{2}N_{c}\delta_{ab}\delta_{kl}
\int\frac{d^{4}q}{(2\pi)^{4}}(g_{\mu\lambda}g_{\nu\sigma}
-g_{\sigma\lambda}g_{\mu\nu})G^{\sigma\lambda}(q)_{kl}. \nonumber \\
\left[\Pi^{ab}_{FP;\mu\nu}(p)\right]_{kl}&=&(k \times l)ig^{2}\delta_{ab}N_{c}
\cdot
\int\frac{d^{4}q}{(2\pi)^{4}}\left[G^{vac}(q)\right]_{kl}
\left[G^{vac}(p-q)\right]_{kl}
\cdot q_{\mu}(q-p)_{\nu}. 
\label{general}
\end{eqnarray}
It has to be remembered that $k,l(=(+,-))$ are not contracted in the 
right hand side of the above equations. So it is obvious that  
${\Pi^{ab}_{Ta;\mu\nu}(p)}_{kl}=0$, when $k \ne l$.  
In the following calculation we neglect the 
vacuum part and concentrate on the medium part of the self energy.
Note that the distribution function is contained only in the medium
part of the self energy. The divergence of the 
vacuum part of the self energy is absorbed in the redefinition of 
the bare quantities which is well known.                  
Like the Green's function the self energy can be written as \cite{pr,stn,wel}:
\be
\Pi_{\mu\nu}(p)~=~T_{\mu\nu}(p)\Pi^T(p)~+~L_{\mu\nu}(p)\Pi^L(p).
\label{pi}
\ee
Using Eq. (\ref{tmunu}) $\Pi$'s are decomposed as:
\begin{eqnarray}
\Pi^{a;T}(p)&=&\Pi^{a}_{\mu\nu}(p)\cdot \frac{ T^{\mu\nu}(p)}{2},\,\,\,\,\,\,
 a=R,A,S,~~~~~~~~{\rm{and}} \nonumber \\
\Pi^{a;L}(p)&=&\Pi^{a}_{\mu\nu}(p)\cdot L^{\mu\nu}(p),\,\,\,\,\,\,
 a=R,A,S. 
\label{comp}
\end{eqnarray}
Total self energy of the medium part is the sum of gluon loop and tadpole
contributions (see Fig. 1) because the ghost loop contribution is present 
in the vacuum
sector in the frozen ghost formalism. The real and imaginary part
of the total self energy can be written as:
$$Re\Pi^{T,L}(p)_{a}=Re\Pi^{T,L}(p)_{Gl;a}+\Pi^{T,L}(p)_{Ta;a},\,\,\,
\,\,a=R,A,S,$$
$$Im\Pi^{T,L}(p)_{a}=Im\Pi^{T,L}(p)_{Gl;a},\,\,\,
\,\,a=R,A,S.$$ 
One can check that the retarded and advanced self energy are related by: 
\begin{eqnarray}
Re\Pi^{T,L}_{Gl;R}(p)=Re\Pi^{T,L}_{Gl;A}(p);
\,\,\,\,\,Im\Pi^{T,L}_{Gl;R}(p)=-Im\Pi^{T,L}_{Gl;A}(p),  \nonumber \\ 
Re\Pi^{T,L}_{Gl;R}(p)=Re\Pi^{T,L}_{Gl;A}(p);
\,\,\,\,\,Im\Pi^{T,L}_{Gl;R}(p)=-Im\Pi^{T,L}_{Gl;A}(p)~~~~~~{\rm{and}} 
\nonumber \\
\,\,\,\,\,\,\,\,\Pi^{T,L}_{Ta;R}(p)=\Pi^{T,L}_{Ta;A}(p).
\label{relation}
\end{eqnarray}
Furthermore we have $\Pi^{T,L}_{Ta;S}(p)$=0.
Therefore we only need calculate $Re\Pi^{T,L}_{Gl;R}$, $Im\Pi^{T,L}_{Gl;R}$, $\Pi^{T,L}_{Gl;S}$
and $\Pi_{Ta;R}$ which we compute in this paper for any non-equilibrium
gluon distribution function $f(p)$. 
Simplifying Eq. (\ref{general}) and using Eq. (\ref{comp}) the explicit 
expression of retarded self energy of the gluon loop diagram is found to be: 

\begin{figure}[htbp]
\begin{center}
\epsfig{file=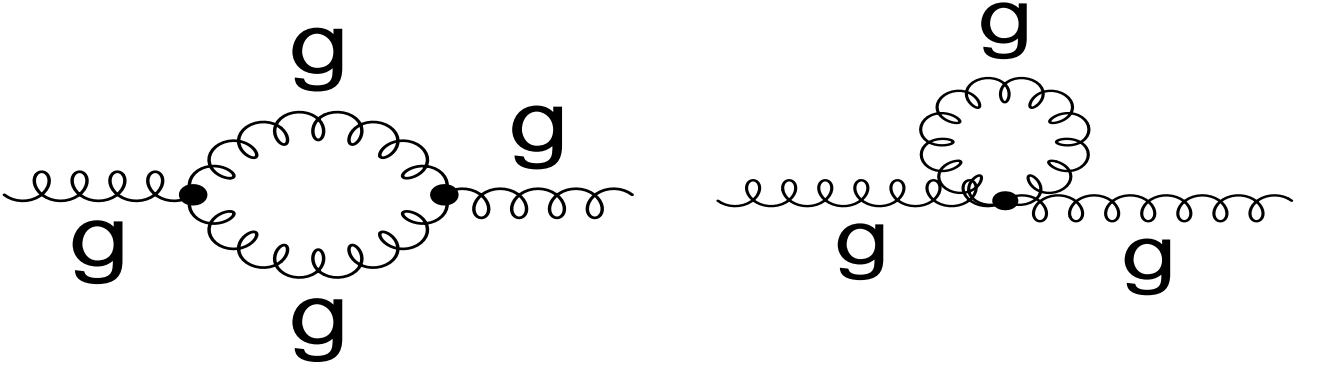}
\caption{Gluon loop and Tadpole diagrams.}
\label{fi1}
\end{center}
\end{figure}

\begin{eqnarray}
\Pi^{T,L}_{Gl;R}(p)&=&\frac{g^{2}}{2}\delta_{ab}N_{c}
\int\frac{d^{4}q}{(2\pi)^{3}}[\frac{\tilde{f}(\vec{q})\delta(q^2)
H^{T,L}(q,p)}{(p-q)^{2}+isgn(p_{0}-q_{0})\epsilon}
\nonumber \\
&+&\frac{\tilde{f}(\vec{p}-\vec{q})\delta((p-q)^2)H^{T,L}(p-q,p)}{q^{2}+isgn(q_{0})\epsilon}],
\end{eqnarray}
where
\begin{eqnarray}
H^{T}(q,p)&=&8\frac{(u\cdot q)(q\cdot p)(u\cdot p)}{(u\cdot p)^{2}-p^2}
-4\frac{(q\cdot p)^{2}}{(u\cdot p)^{2}-p^2} 
-4\frac{p^2(q\cdot u)^{2}}{(u\cdot p)^{2}-p^2} \nonumber \\
&-&\left[(p+q)^{2}\right]\left[1-
\frac{(q\cdot p)(u\cdot p)}{(u\cdot q)((u\cdot p)^{2}-p^2)}
+\frac{(q\cdot p)^{2}}{2(u\cdot q)^{2}((u\cdot p)^{2}-p^2)}+
\frac{p^2}{2((u\cdot p)^{2}-p^2)}\right] \nonumber \\
&-&4p^{2}+8\frac{(q\cdot p)(u\cdot p)}{(u\cdot q)}-4
\frac{(q\cdot p)^{2}}{(u\cdot q)^{2}} \nonumber \\
&+&\frac{(\xi-1)(p^4(- (q \cdot p)^2+2 (q \cdot u) (p \cdot u) (q \cdot p) 
+(q \cdot u)^2 (p^2-2 (p \cdot u)^2)))}
{2(q \cdot u)^2( (q \cdot p)-p^2)(p^2- (p \cdot u)^2)}
\end{eqnarray}

and

\begin{eqnarray}
H^{L}(q,p)&=&\frac{8p^{2}}{(u\cdot p)^{2}-p^{2}}[((u\cdot q)
-\frac{(u\cdot p)(q\cdot p)}{p^2})^2] \nonumber \\
&-&\left[(p+q)^{2}\right] 
\left[\frac{2(q\cdot p)(p\cdot u)^3}{(q \cdot u)p^2((p \cdot u)^2-p^2)}
-\frac{(q\cdot p)^2(p\cdot u)^2}{(q \cdot u)^2p^2((p \cdot u)^2-p^2)}
-\frac{(p\cdot u)^2}{((p \cdot u)^2-p^2)}
\right] \nonumber \\
&-&4p^{2}+8\frac{(q\cdot p)(u\cdot p)}{(u\cdot q)}-4
\frac{(q\cdot p)^{2}}{(u\cdot q)^{2}} \nonumber \\
&+& (\xi-1)\frac{(p \cdot u)^2 ( (q \cdot p)^2-2 (q \cdot u) 
(p \cdot u) (q \cdot p)+ (q \cdot u)^2p^2)}{(q \cdot u)^2
(p^2- (p \cdot u)^2)}.
\end{eqnarray}

Using the $\delta$ function we get:
\begin{eqnarray}
\Pi^{T,L}_{Gl;R}(p)&=&\frac{g^{2}}{2}\delta_{ab}N_{c}
\int\frac{d^{3}q}{(2\pi)^{3}}\frac{1}{2|\vec{q}|} 
[\frac{f(\vec{q})\cdot H^{T,L}(q,p)|_{q_0=|\vec{q}|}}{(p_{0}-|\vec{q}|)^{2}-|\vec{q}
-\vec{p}|^{2}+isgn(p_{0}-|\vec{q}|)\epsilon}
\nonumber \\
&+&\frac{f(-\vec{q})\cdot H^{T,L}(q,p)|_{q_0=-|\vec{q}|}}{(p_{0}+|\vec{q}|)^{2}-|\vec{q}
-\vec{p}|^{2}+isgn(p_{0}+|\vec{q}|)\epsilon}]
\nonumber \\
&+&(q\rightarrow p-q).
\label{pitl}
\end{eqnarray}
Similar but more complicated calculation for symmetric self energy yields:
\begin{eqnarray}
\Pi^{(T,L)}_{Gl;S}(p)&=&\frac{g^{2}}{2i}\delta_{ab}N_{c}
\int\frac{d^{3}q}{(2\pi)^{2}}\delta\left((|\vec{q}|-p_{0})^{2}
-|\vec{q}-\vec{p}|^{2}\right)\cdot \frac{K^{(T,L)}(q,p)|_{q_{0}=
|\vec{q}|}}{2|\vec{q}|} \cdot f(\vec{q})
\cdot \tilde{f}(\vec{p}-\vec{q}) \nonumber \\
&+&\delta\left((|\vec{q}|+p_{0})^{2}
-|\vec{q}-\vec{p}|^{2}\right)\cdot \frac{K^{(T,L)}(q,p)|_{q_{0}=
-|\vec{q}|}}{2|\vec{q}|} \cdot f(-\vec{q})
\cdot \tilde{f}(\vec{p}-\vec{q}),
\label{pits}
\end{eqnarray}
where $K^{T,L}(q,p)$ are given by:
\begin{eqnarray}
&&K^T(q,p)=\frac{p^2(p^2+4 (q\cdot u)((q \cdot u)-(p\cdot u)))}{4(q\cdot u)^2(q\cdot u-p\cdot u)^2(p^2-(p\cdot u)^2)}
[8(q\cdot u)^4-16(p\cdot u)(q\cdot u)^3 \nonumber \\
&-&12(p^2-2(p\cdot u)^2)  
\cdot(q\cdot u)^2+4(3p^2(p\cdot u)-4(p \cdot u)^3)(q\cdot u)+8(p\cdot u)^4+p^4-8p^2
(p\cdot u)^2]
\end{eqnarray}
and
\vspace{0.3cm}
\begin{eqnarray}
K^L(q,p)&=& \frac{p^2[(p \cdot u)-2 (q \cdot u)]^2}{4(q \cdot u)^2(q \cdot u-p \cdot u)^2(p^2- (p \cdot u)^2)}
[8 (q \cdot u)^4-16 (p \cdot u) (q \cdot u)^3 \nonumber \\
&+&4(2 (p \cdot u)^2+p^2) (q \cdot u)^2-4p^2 (p \cdot u) (q \cdot u)+p^4]. 
\end{eqnarray}
The expressions for the transverse and longitudinal part of the
self energy coming from the
Tadpole part (Eq. (\ref{general}) and (\ref{comp})) are found to be:
\begin{small}
\begin{eqnarray}
&&\Pi^{T}_{Ta;R}(p)=g^{2}\delta_{ab}N_{c}
\int\frac{d^{3}q}{(2\pi)^{3}}\frac{f(\vec{q})}{2|\vec{q}|} 
[1
+\frac{(u\cdot p)(q\cdot p)}{2(u\cdot q)((u\cdot p)^{2}-p^{2})}-\frac{(q\cdot p)^{2}}
{2(u\cdot q)^{2}((u\cdot p)^{2}-p^{2})}
\nonumber \\
&-&\frac{p^{2}}{2((u\cdot p)^{2}-p^{2})}
]|_{q_{0}=|\vec{q}|}
+\frac{f(-\vec{q})}{2|\vec{q}|}[1
+\frac{(u\cdot p)(q\cdot p)}{2(u\cdot q)((u\cdot p)^{2}-p^{2})}-\frac{(q\cdot p)^{2}}
{2(u\cdot q)^{2}((u\cdot p)^{2}-p^{2})} \nonumber \\
&-&\frac{p^{2}}{2((u\cdot p)^{2}-p^{2})}]|_{q_{0}=-|\vec{q}|}
\label{pita}
\end{eqnarray}
\end{small}
and
\begin{small}
\begin{eqnarray}
&&\Pi^{L}_{Ta;R}(p)={g^{2}}\delta_{ab}N_{c}
\int\frac{d^{3}q}{(2\pi)^{3}}\frac{f(\vec{q})}{2|\vec{q}|} 
[3-2\frac{(u\cdot p)(q\cdot p)}{p^{2}(u\cdot q)}+\frac{p^{2}}{(u\cdot p)^{2}-p^{2}}(1-\frac{(u\cdot p)(q\cdot p)}{p^{2}(u\cdot q)})^{2}]|_{q_{0}=|\vec{q}|}
\nonumber \\
&+&\frac{f(-\vec{q})}{2|\vec{q}|}[3-2\frac{(u\cdot p)(q\cdot p)}{p^{2}(u\cdot q)}+\frac{p^{2}}{(u\cdot p)^{2}-p^{2}}(1-\frac{(u\cdot p)(q\cdot p)}{p^{2}(u\cdot q)})^{2}]|_{q_{0}=-|\vec{q}|}
\label{pila}
\end{eqnarray}
\end{small}
respectively. To summarize Eqs. 
(\ref{pitl}), (\ref{pits}), (\ref{pita}) and (\ref{pila}) 
contain all the expressions for
different parts of the self energy in non-equilibrium which will be
used in the medium modified propagator to obtain a finite collision
term. In this paper we make hard momentum loop approximation in non-equilibrium
which is simillar to hard thermal loop approximation in equilibrium \cite{stn}.
In the hard momentum loop approximation the self energies are found to be
gauge parameter independent (see Appendix).   

To obtain the collision term for the process 
$gg \rightarrow gg$ in the medium we have to use the resummed 
Feynman propagator $[\tilde{G}(p)_{++}]$
which can be obtained from the resummed advanced,
retarded and symmetric propagators via the relation:
\be
\left[\tilde{G}(p)\right]_{++}=\frac{1}{2}
\left[\tilde{G}_{S}(p)+\tilde{G}_{A}(p)+\tilde{G}_{R}(p)\right], 
\ee
where 
\begin{equation}
\tilde{G}_{R,A}(p)=\frac{1}{p^2-\Pi_{R,A}(p) \pm i sgn(p_{0})\epsilon},
\end{equation}
and
\begin{eqnarray}
\tilde{G}_{S}(p)&=&\left[1+2f(\vec{p})\right]sgn(p_{0})[\tilde{G}_{R}(p)
-\tilde{G}_{A}(p)] \nonumber \\
&+&\left(\Pi_{S}(p)-(1+2f(\vec{p}))sgn(p_{0})[\Pi_{R}(p)-\Pi_{A}(p)]\right)
\times \tilde{G}_{R}(p)\times \tilde{G}_{A}(p).
\end{eqnarray}
Different parts of the self energy appearing in the above 
equations are given in Eqs. 
(\ref{pitl}), (\ref{pits}), (\ref{pita}) and (\ref{pila}). 
Finally, using the relation between various parts of the self energy 
(see Eq. (\ref{relation})) we find from the above equations:
\begin{equation}
\tilde{G}_{++}(p)=\frac{p^2-Re\Pi_{R}(p)+\frac{1}{2}\Pi_{S}(p)}{(p^2-Re\Pi_{R}(p))^2+(Im\Pi_{R}(p))^2}
\end{equation}
which is the required expression for the medium modified resummed Feynman
gluon propagator in non-equilibrium at one loop level of the self energy.

Our main purpose is to remove infrared divergence appearing in the
small angle partonic scatterings which plays an important 
role in the production and equilibration of minijet plasma.  
For this purpose it is necessary to study the infrared behaviour
of the self energies obtained in this paper for non-equilibrium
situtations. We note that in the static limit ($p_0=0, |\vec p| \rightarrow 0$)
one obtains $\Pi^L(p_0=0, |\vec p| \rightarrow 0)= m_D^2$ (the
Debye screening mass) and
$\Pi^T(p_0=0, |\vec p| \rightarrow 0)= m_g^2$ (the
magnetic screening mass). It is widely believed that while Debye screening
mass is non-zero the magnetic screening mass is zero at one loop
level of the self energy. For this reason one expects that while 
the electric field is screened, the magnetic field is not screened and
one still has infrared divergence in the magnetic sector
at one loop level. However, this is true for a system where the
gluon distribution function is isotropic (in momentum space)
or in equilibrium. This is not true for non-isotropic gluon
distribution function which is the case in the early stage
of the heavy-ion collisions at RHIC and LHC. The static limit
result of the transverse part of the self energy derived in this 
paper is not zero for non-isotropic (in momentum space)
gluon distribution function.
The typical values of the magnetic screening masses at one loop
level is found to be 257 MeV at RHIC and 330 MeV at LHC by using
non-equilibrium gluon-minijet distribution function \cite{fred}
at the initial time. 
Hence the medium modified gluon propagator derived in this paper
is safe from infrared divergences both in electric and magnetic
sector in non-equilibrium situations at RHIC and LHC. 
Therefore these propagators
can be used to obtain finite collision terms to study equilibration
of minijet plasma at RHIC and LHC. Note that the above values are
obtained by using the minijet distribution function at the initial
time. The time evolution of these screening masses have to be
determined by solving relativistic transport equations with
Bjorken's boost invariance picture taken into account. As these
involves extensive numerical work (see \cite{nayak}) we will 
report the self-consistent space-time evolution study elsewhere.

\section{Conclusions}
In this paper we have applied the closed-time
path integral formalism to study medium effects of the minijet plasma
in non-equilibrium. In particular we have derived the medium 
modified resummed gluon propagator to the one loop order of self
energy which is
necessary to obtain a finite collision term
to study equilibration of parton plasma at RHIC and LHC. 
These medium
modified propagators have been studied in more detail in finite temperature
QCD formulations. However, at RHIC and LHC the parton momentum distributions
at early stage
are anisotropic and finite temperature QCD formulations can not be
applied to these non-equilibrium situations.
This is
because a parton inside the nucleus (which travels almost at the speed of light
at RHIC and LHC) carries mostly longitudinal momentum before an ultra
relativistic nuclear collisions. After jets and minijets are formed and 
suffer secondary collisions the isotropy between longitudinal 
and transverse momentum may be achieved. 
According to Bjorken's proposal \cite{bjorken} many quantities 
are expected to be expressed in terms of boost invariant parameters. 
For this reason and to have a covariant formulation we have worked in the
covariant gauge which is a suitable gauge for expanding plasma.
We give the result of the resummed gluon propagator upto one loop level
of self energy in non-equilibrium. Furthermore it is shown that
these propagators are infrared divergence free, both in electric and
magnetic sector.

The medium modified resummed gluon propagator we derived in this
paper will be used to obtain finite collision term 
for the $gg \rightarrow gg$ scattering process in
non-equilibrium to solve the relativistic kinetic equation (Eq. (\ref{trs1}))
to study equilibration of minijet plasma at RHIC and LHC. 
In this way one does not have to put
ad-hoc values for the momentum transfer cut-off 
which crucially changes all the properties 
and hence determination of all the signatures of the quark-gluon 
plasma. 
In future we hope to use our results of the resummed gluon propagators
in the collision term to study production and equilibration of the
minijet plasma at RHIC and LHC.
As the solution of the relativistic transport equation involves extensive
numerical work (see \cite{nayak}) we will report it elsewhere.

\acknowledgements
We thank Fred Cooper for useful discussions. We thank Joerg Ruppert for his
help in using Feynman package. C-W. K. and G. C. N. acknowledge the financial
support from Alexander von Humboldt Foundation.

\pagebreak
{\bf Appendix A: The explicit forms of self energies in Hard Loop Momentum Approximation:}\\

We use hard momentum loop (HML) approximation in non-equilibrium
which is equivalent to hard thermal loop (HTL) approximation in
equilibrium. In the hard momentum loop approximation the  
loop momentum $q$ is harder than external momentum $p$ \cite{stn}. In HML 
approximation we find that
the whole expressions are independent of gauge-fix parameters $\xi$. 
In the hard momentum loop approximation we find:
\begin{eqnarray}
Re\Pi^{T}_{Gl;R}(p)&\sim &\frac{g^{2}}{2}\delta_{ab}N_{c}
\frac{1}{(u\cdot p)^2-p^2}
\int\frac{d^{3}q}{(2\pi)^{3}}\frac{1}{2|\vec{q}|} 
[\frac{f(\vec{q})}{(|\vec{q}|-p_{0})^{2}-|\vec{q}
-\vec{p}|^{2}} \nonumber \\
&\cdot&[8(u\cdot q)(q\cdot p)(u\cdot p)
-4(q\cdot p)^{2} 
-4p^2(q\cdot u)^{2} \nonumber \\
&-&2(p\cdot q)(u\cdot p)^2+p^2(p\cdot q)+2\frac{(p\cdot q)^{2}(u\cdot p)}{u\cdot q}
-2\frac{(q\cdot p)^3}{(u\cdot q)^2}]|_{q_0=|\vec{q}|}
\nonumber \\
&+&\frac{f(-\vec{q})}{(|\vec{q}|+p_{0})^{2}-|\vec{q}
-\vec{p}|^{2}} 
[8(u\cdot q)(q\cdot p)(u\cdot p)
-4(q\cdot p)^{2} 
-4p^2(q\cdot u)^{2} \nonumber \\
&-&2(p\cdot q)(u\cdot p)^2+p^2(p\cdot q)+2\frac{(p\cdot q)^{2}(u\cdot p)}{u\cdot q}
-2\frac{(q\cdot p)^3}{(u\cdot q)^2}]
|_{q_0=-|\vec{q}|}]
\nonumber \\
&+&(q\rightarrow p-q),  
\end{eqnarray}
and
\begin{eqnarray}
Im\Pi^{T}_{Gl;R}(p)&\sim &\frac{g^{2}}{2}\delta_{ab}N_{c}
\frac{1}{(u\cdot p)^2-p^2}
\int\frac{d^{3}q}{(2\pi)^{2}}
sgn(p_{0}-|\vec{q}|)
\delta\left((p_{0}-|\vec{q}|)^{2}
-|\vec{q}-\vec{p}|^{2}\right) 
\frac{f(\vec{q})}{2|\vec{q}|} 
\nonumber \\
&&[8(u\cdot q)(q\cdot p)(u\cdot p)
-4(q\cdot p)^{2} 
-4p^2(q\cdot u)^{2} \nonumber \\
&-&2(p\cdot q)(u\cdot p)^2+p^2(p\cdot q)+2\frac{(p\cdot q)^2(u\cdot p)}{u\cdot q}
-2\frac{(q\cdot p)^3}{(u\cdot q)^2}
]|_{q_0=|\vec{q}|} \nonumber \\
&+& sgn(p_{0}+|\vec{q}|)
\delta\left((|\vec{q}|+p_{0})^{2}
-|\vec{q}-\vec{p}|^{2}\right) 
\frac{f(-\vec{q})}{2|\vec{q}|} 
\nonumber \\
&&[8(u\cdot q)(q\cdot p)(u\cdot p)
-4(q\cdot p)^{2} 
-4p^2(q\cdot u)^{2} \nonumber \\
&-&2(p\cdot q)(u\cdot p)^2+p^2(p\cdot q)+2\frac{(p\cdot q)^2(u\cdot p)}{u\cdot q}
-2\frac{(q\cdot p)^3}{(u\cdot q)^2}]|_{q_0=-|\vec{q}|}
\nonumber \\
&+&(q\rightarrow p-q), 
\end{eqnarray}
Similarly for longitudinal components we obtain:
\begin{eqnarray}
Re\Pi^{L}_{Gl;R}(p)&\sim &\frac{g^{2}}{2}\delta_{ab}N_{c}\frac{1}{(u\cdot p)^{2}-p^{2}}
[\int\frac{d^{3}q}{(2\pi)^{3}}\frac{f(\vec{q})}{2|\vec{q}|} 
\frac{1}{(p_{0}-|\vec{q}|)^{2}-|\vec{q}
-\vec{p}|^{2}}[8p^2((u\cdot q)
-\frac{(u\cdot p)(q\cdot p)}{p^2})^2 \nonumber \\
&-&4\frac{(p\cdot q)^2(u\cdot p)^3}{(u\cdot q)p^2}+2\frac{(q\cdot p)^3(u\cdot p)^2}{(u\cdot q)^2p^2}
+2(u\cdot p)^2(q\cdot p)]
_{q_{0}=|\vec{q}|} \nonumber \\
&+&\int\frac{d^{3}q}{(2\pi)^{3}}\frac{f(-\vec{q})}{2|\vec{q}|} 
\frac{1}{(p_{0}+|\vec{q}|)^{2}-|\vec{q}
-\vec{p}|^{2}} 
[8p^2((u\cdot q)
-\frac{(u\cdot p)(q\cdot p)}{p^2})^2 \nonumber \\
&-&4\frac{(p\cdot q)^2(u\cdot p)^3}{(u\cdot q)p^2}+2\frac{(q\cdot p)^3(u\cdot p)^2}{(u\cdot q)^2p^2}
+2(u\cdot p)^2(q\cdot p)]
_{q_{0}=-|\vec{q}|}] 
\nonumber \\
&+&(q\rightarrow p-q),
\end{eqnarray}
and
\begin{eqnarray}
Im\Pi^{L}_{Gl;R}(p)&\sim &\frac{g^{2}}{2}\delta_{ab}N_{c}\frac{8p^{2}}{(u\cdot p)^{2}-p^{2}}
\int\frac{d^{3}q}{(2\pi)^{2}}\frac{f(\vec{q})}{2|\vec{q}|} \nonumber \\ 
&&sgn(p_{0}-|\vec{q}|)
\delta\left((|\vec{q}|-p_{0})^{2}
-|\vec{q}-\vec{p}|^{2}\right)
[8p^2((u\cdot q)
-\frac{(u\cdot p)(q\cdot p)}{p^2})^2) \nonumber \\
&-&4\frac{(p\cdot q)^2(u\cdot p)^3}{(u\cdot q)p^2}+2\frac{(q\cdot p)^3(u\cdot p)^2}{(u\cdot q)^2p^2}
+2(u\cdot p)^2(q\cdot p)]
_{q_{0}=|\vec{q}|}] \nonumber \\
&+&\frac{f(-\vec{q})}{2|\vec{q}|}sgn(p_{0}+|\vec{q}|)
\delta\left((|\vec{q}|+p_{0})^{2}
-|\vec{q}-\vec{p}|^{2}\right)
[8p^2((u\cdot q)
-\frac{(u\cdot p)(q\cdot p)}{p^2})^2) \nonumber \\
&-&4\frac{(p\cdot q)^2(u\cdot p)^3}{(u\cdot q)p^2}+2\frac{(q\cdot p)^3(u\cdot p)^2}{(u\cdot q)^2p^2}
+2(u\cdot p)^2(q\cdot p)]
_{q_{0}=-|\vec{q}|}] \nonumber \\
&+&(q\rightarrow p-q)
\end{eqnarray}
For the symmetric part of the self energy we find:
\begin{eqnarray}
\Pi^{(T)}_{Gl;S}(p)&=&\frac{g^{2}}{2i}\delta_{ab}N_{c}\frac{4p^2}{(u\cdot p)^2-p^2}
\int\frac{d^{3}q}{(2\pi)^{2}}\delta\left((|\vec{q}|-p_{0})^{2}
-|\vec{q}-\vec{p}|^{2}\right)\cdot|\vec{q}|(u_0-\vec{u}\cdot\hat{q})^2
f(\vec{q})\cdot \tilde{f}(\vec{p}-\vec{q}) \nonumber \\
&+&\delta\left((|\vec{q}|+p_{0})^{2}
-|\vec{q}-\vec{p}|^{2}\right)\cdot|\vec{q}|(-u_0-\vec{u}\cdot\hat{q})^2 
f(-\vec{q})
\cdot \tilde{f}(\vec{p}-\vec{q}).
\end{eqnarray}
It is interesting to see that in the Hard Loop Momentum Approximation we have
$\Pi^{(T)}_{Gl;S}(p)\sim \Pi^{(L)}_{Gl;S}(p)$.


\begin{thebibliography}{aaaaaaaaaaaaaaa}
\bibitem{nayak} G. C. Nayak, A. Dumitru, L. McLerran and W. Greiner, 
Nucl. Phys. A687 (2001) 457. 
\bibitem{keijo1} K. J. Eskola and K. Kajantie, Z. Phys. C. 75 (1997) 515;
N. Hammon, H. St\"ocker and W. Greiner, Phys.
Phys. Rev. C61, 014901 (2000);
A. Krasnitz and R. Venugopalan, Phys. Rev. Lett. 84 (2000) 4309.
\bibitem{satz} T. Matsui and H. Satz, \textit{Phys. Lett.} \textbf{B
178} (1986) 416; G. C. Nayak, JHEP 9802 (1998) 005.
\bibitem{str} J. Rafelski and B. Mueller, \textit{Phys. Rev. Lett.}
\textbf{48} (1981) 1066
\bibitem{dil} M.T. Strickland, \textit{Phys. Lett.} \textbf{B 331}
(1994) 245; Gouranga C. Nayak, \textit{Phys. Lett.} \textbf{B 442}
(1995) 427
\bibitem{ph} E.V. Shuryak and L- Xiong, \textit{Phys. Rev. Lett.}
\textbf{70} (1993) 2274
\bibitem{geiger} K. Geiger, Phys. Rep. 280 (1995) 237;
K. Geiger and J. I. Kapusta, Phys. Rev. D47, (1993) 4905.
\bibitem{wang12}
H. Heiselberg, X.N. Wang, Nucl. Phys. B 462:389-414 (1996). 
\bibitem{wong}
S.H.M. Wong, Phys. Rev C54 (1996) 2588; Phys. Rev. C56 (1997) 1075.
\bibitem{gyulassy}
M. Gyulassy, Y. Pang, B. Zhang, Nucl. Phys. A 626 (1997) 999;
B. Zhang, Comput. Phys. Commun. 109 (1998) 193.
\bibitem{bhal}
R.S. Bhalerao, G.C. Nayak, Phys. Rev. C 61 (2000) 054907.
\bibitem{naya} G. C. Nayak and V. Ravishankar, Phys. Rev D55 (1997) 6877;
Phys. Rev. C58 (1998) 356.
\bibitem{others}
Y. Kluger, J.M. Eisenberg, B. Svetitsky, F. Cooper and E. Motolla,
Phys. Rev. Lett. 67 (1991) 2427; F. Cooper, J. M. Eisenberg, Y.
Kluger, E. Motolla and B. Svetitsky, Phys. Rev. D 48 (1993) 190;
J.M. Eisenberg Phys. Rev. D 36 (1987) 3114; {\it ibid} D 40 (1989)
456; T. S. Biro, H.B. Nielsen and J. Knoll, Nucl Phys. B 245
(1984) 449; M. Herrmann and J. Knoll, Phys. Lett. B 234 (1990)
437; D. Boyanovsky H.J. de Vega, R. Holman, D.s. Lee and A. Singh,
Phys. Rev. D 51 (1995) 4419; H. Gies Phys. Rev. D 61 (2000)
085021.
\bibitem{denis} D. D. Dietrich, G. C. Nayak and W. Greiner, Phys. Rev. D64
(2001) 074006; hep-ph/0009178; hep-ph/0202144, 
J. Phys. G (in press); 
G.C. Nayak, D. D. Dietrich and W. Greiner, Rostock 2000/Trento
2001, Exploring Quark Matter 71-78, hep-ph/0104030.
\bibitem{larry1} L. Mclerran and R. Venugopalan, Phys. Rev. D 49 (1994) 2233,
Phys. Rev. D 49 (1994) 3352; Yu. Kovchegov and A. H. Mueller, Nucl. Phys.
B 529 (1998) 451; A. Kovner, L. McLerran and H. Weigert, Phys. Rev D
52 (1998) 3809, Phys. Rev. D 52 (1998) 6231; Y. V. Kovchegov, E. Levin and L. 
McLerran, hepph/9912367.
\bibitem{roberts}
C. D. Roberts and S. M. Schmidt, Prog. Part. Nucl. Phys. 45
Suppl.1:1-103, 2000; V. Vinik, {\it et. al}, Eur. Phys. J.C22 (2001) 341; 
J.C.R. Bloch, C. D. Roberts and S. M. Schmidt, Phys. Rev. D61 (2000) 117502;
J. C. R. Bloch, {\it et. al}, Phys. Rev. D (1999) 116011.
\bibitem{bla} J-P. Blaizot and E. Iancu, Nucl. Phys.
B570 (2000) 326; Nucl. Phys. B417 (1994) 608.
\bibitem{art} R. D. Pisarski, Phys. Rev. D47 (1993) 5589; T. Altherr,
Phys. Lett. B341 (1995) 325; R. Baier, M. Dirks, K. Redlich and
D. Schiff, Phys. Rev. D56 (1997) 2548; and references therein.
\bibitem{naka} Y. Yamanaka, {\it et. al}, Int. J. Mod. Phys. A9 (1994) 
1153; A. Niegawa, Phys. Rev. D65 (2002) 056009, and references therein.
\bibitem{bll2} J-P. Blaizot and E. Iancu, Phys. Rept. 359 (2002) 355 and
references therein. 
\bibitem{ruppert}
J. Ruppert, G. C. Nayak, D. D. Dietrich, H. Stoecker and 
W. Greiner, Phys. Lett. B520 (2001)
233.
\bibitem{sch} J. Shwinger, J. Math. Phys. 2 (1961) 407.
\bibitem{mah} K. T. Mahanthappa, Phys. Rev. 126 (1962) 329; P. M. Bakshi and
K. T. Mahanthappa, J. Math. Phys. 4 (1963) 1; 
L. V. Keldysh, Sov. Phys. JETP 20 (1965) 1018.
\bibitem{otrs}L. P. Kadanoff and G. Baym, {\it Quantum Statistical
Mechanics} (Benjamin, New York, 1962); K. Chou, Z. Su, B. Hao and
L. Yu, Phys. Rept. 118 (1985) 1; F. Cooper, hep-th/9504073 and references
therein. 
\bibitem{hu} E. Calzetta and B. L. Hu, Phys. Rev. D 37 (1988) 2878.
\bibitem{bjorken} J. D. Bjorken, Phys. Rev. D. 27 (1983) 140.
\bibitem{fred} 
F. Cooper, C-W. Kao and G. C. Nayak, hep-ph/0204042.
\bibitem{muller} A. H. Mueller, Nucl. Phys. B572 (2000) 227; 
Phys. Lett.B475 (2000) 220; R. Baier, A. H. Mueller,
D. Schiff and D. T. Son, Phys. Lett. B502 (2001) 51.
\bibitem{land} P. V. Landshoff and A. Rebhan, Nucl. Phys. B 383 (1992)
607 and Erratum, {\it ibid} 406 (1993) 517.
\bibitem{petro} M. D'Attanasio and M. Pietroni, Nucl. Phys. B 498 (1997) 443.
\bibitem{pr} N. P. Landsman and Ch.G. van Weert, Phys. Rep. 145 (1987) 141.
\bibitem{wel} H. A. Weldon, Phys. Rev. D 26 (1982) 1394.
\bibitem{stn} M. H. Thoma, hep-ph/0010164; 
S. Mrowczynski and M. H. Thoma, Phys. Rev. D 62 (2000) 036011.

\end{thebibliography}
\end{document}